\documentclass[12pt]{article}

\usepackage{latexsym}
\usepackage{graphicx}
\usepackage{booktabs}
\usepackage{amsmath}
\usepackage{authblk}

\usepackage{url}
\usepackage[top=1in,bottom=1in,left=1in,right=1in]{geometry}
\usepackage{float}

\newcommand{\rateMatrix}{\ensuremath{\vec{\Lambda}}}
\newcommand{\rateMatrixSpace}{\ensuremath{\vec{\Lambda}} }
\newcommand{\rateMatrixElement}{\ensuremath{\Lambda_{ij}}}
\newcommand{\seats}{\ensuremath{\Psi_{ij}}}
\newcommand{\coefficient}{\ensuremath{\beta}}
\newcommand{\coefficientSpace}{\ensuremath{\beta} }
\newcommand{\indicator}{\ensuremath{\delta}}
\newcommand{\indicatorSpace}{\ensuremath{\delta} }
\newcommand{\predictor}{\textit{p}}

\usepackage{ctable}

\begin{document}

\title{The seasonal flight of influenza: a unified framework for spatiotemporal hypothesis testing}

\author[1]{Philippe Lemey}
\author[2,3]{Andrew Rambaut}
\author[2,4,5]{Trevor Bedford}
\author[1]{Nuno Faria}
\author[1]{Filip Bielejec}
\author[1]{Guy Baele}
\author[3,6,7]{Colin A. Russell}
\author[3,6,7,8]{Derek J. Smith}
\author[9]{Oliver G. Pybus}
\author[10,11]{Dirk Brockmann}
\author[12,13]{Marc A. Suchard}

\affil[1]{Department of Microbiology and Immunology, KU Leuven, Leuven, Belgium.}
\affil[2]{Institute of Evolutionary Biology, University of Edinburgh, Edinburgh, UK.}
\affil[3]{Fogarty International Center, National Institutes of Health, Bethesda, MD, USA.}
\affil[4]{Department of Ecology and Evolutionary Biology, University of Michigan, Ann Arbor, Michigan, USA.} 
\affil[5]{Howard Hughes Medical Institute, University of Michigan, Ann Arbor, Michigan, USA.}
\affil[6]{Department of Zoology, University of Cambridge, Cambridge, UK.}
\affil[7]{World Health Organization (WHO) Collaborating Center for Modeling, Evolution, and Control of Emerging Infectious Diseases, Cambridge, UK.}
\affil[8]{Department of Virology, Erasmus Medical Centre, Rotterdam, Netherlands.}
\affil[9]{Department of Zoology, University of Oxford, Oxford
, United Kingdom.}
\affil[10]{Engineering Sciences and Applied Mathematics, Northwestern University, 
Evanston IL
, USA.}
\affil[11]{Northwestern Institute on Complex Systems, Evanston, Illinois, USA.}
\affil[12]{Departments of Biomathematics and Human Genetics, David Geffen School of Medicine, University of California, Los Angeles, CA
, USA.}
\affil[13]{Department of Biostatistics, School of Public Health, University of California, Los Angeles, CA
, USA.}

\maketitle

\newpage

\section{Abstract}
Global mobility flow data are at the heart of spatial epidemiological models used to predict infectious disease behavior but this wealth of data on human mobility has been largely neglected by reconstructions of pathogen evolutionary dynamics using viral genetic data.
Although stochastic models of viral evolution may potentially be informed by such data, a major challenge lies in deciding which mobility processes are critical and to what extent they contribute to shaping contemporaneous distributions of pathogen diversity.
Here, we develop a framework to integrate predictors of viral diffusion with phylogeographic inference and estimate human influenza H3N2 migration history while simultaneously testing and quantifying the factors that underly it. We provide evidence for air travel governing the global dynamics of human influenza whereas other processes act at a more local scale.

\vspace{10 mm}

\section{Introduction}

Global public health is repeatedly and increasingly challenged by the emergence of high-impact pathogens \cite{jones08}.
Novel influenza strains, severe acute respiratory syndrome (SARS) virus and Methicillin-resistant \textit{Staphylococcus aureus} represent only a few examples of pathogens that exploited today's complex and voluminous human traffic and mobility to rapidly disseminate in our globalized world.

The worldwide air transportation network is by far the most extensively studied mobility system in the context of human infectious disease dynamics \cite{brockmann09}.
Indeed, air travel represents an obvious driving force for the global circulation of seasonal influenza A (H3N2) viruses, and may explain the absence of locally persistent strains in between epidemic seasons \cite{russell08}.
Retrospective modeling of the 1968 `Hong Kong flu' pandemic spread demonstrated that the H3N2 virus diffused through a network of global cities interconnected by air travel \cite{rvachev85}.
Numerous modeling studies have subsequently examined the influence of air travel on influenza spread (e.g. \cite{longini86,flahaut88,grais03,grais04}),
but far less work has attempted to verify such models against underlying patterns of host movement \cite{viboud06}.

Two studies on the timing and rate of seasonal influenza spread across the United States have highlighted the difficulty of resorting to 
standard epidemiological
data to disentangle the contributions of different human transportation systems to influenza spread.
Using weekly time series of excess mortalities between 1972 and 2002, Viboud et al. (2006) \cite{viboud06} demonstrated that the patterns of timing and incidence across the continental United States are significantly associated with euclidean distance and various measures of domestic transportation (including airline travel), but most strongly with rates of movement of people to and from their workplaces.
In the same year, Brownstein et al. (2006) \cite{brownstein06} demonstrated that the rate of inter-regional spread and timing of influenza in the United States, as measured using weekly influenza and pneumonia mortality statistics from 1996 to 2005, is predicted by domestic airline travel volume in November.
Because both studies implicated a different key driver of seasonal influenza spread across the United States, the findings were subject of debate \cite{viboud06b}, in particular in the light of a mounting threat of an influenza pandemic \cite{monto05,webby03} and the need for decisions on implementing travel restrictions.
Insights at the global scale are also revealed by simulation studies (e.g, \cite{balcan09}), and empirical analysis using global mortality data may prove even more challenging.

As a historical record of epidemic spread, viral genetic data may offer a valuable alternative for empirical verification of epidemic models.
The power of fitting statistical models of evolution to observed sequence data has clearly been demonstrated by a number of seminal studies, e.g. by revealing the genetic dynamics of influenza A H3N2 seasonality \cite{rambaut08} and 
spatial patterns of global H3N2 circulation \cite{russell08,bedford10}.
More generally, viral phylogenetic and epidemiological insights have culminated into a phylodynamic framework that unifies evolutionary and ecological dynamics to explain patterns of viral diversity \cite{grenfell04}.
Although model-based inference is increasingly used to reconstruct viral diffusion through time and space, these attempts typically fit parameter-rich models to sparse spatial data, and post-hoc interpretation of phylogeographic patterns are then difficult to relate directly to underlying ecological and evolutionary processes \cite{holmes09}.

Here, we present a novel 
model-based approach to simultaneously reconstruct spatio-temporal history and test the contribution of potential diffusion predictors.
Our phylogeographic reconstruction considers discrete sampling locations defined by geographical and administrative boundaries as well as air communities identified through direct analysis of the global air transportation network.
By parameterizing the discrete diffusion process in terms of the inclusion and contribution of predictors, our approach generally requires considerably fewer parameters compared to standard phylogeographic inference.
We demonstrate how this model allows for the integration of viral genetic data and human mobility measures to draw inference about key drivers of global influenza dynamics.

\section{Methods}

\subsection{Sequence and location data}\label{genetic}

We compiled 1441 hemagglutinin sequences with known date and location of sampling previously obtained by
\cite{russell08}.
These sequences were sampled globally from 2002 to 2007 and are representative of a larger sampling (13,000 isolates) used for antigenic analysis \cite{russell08}.
We explored different spatial and air travel-assisted subdivisions with subsampling to examine the impact of discrete sampling allocation and sample numbers per locations on our phylogeographic estimates.
In an attempt to include all sequence data while keeping the number of samples per location as balanced as possible, we first divided all the sequences into 26 geographic regions (Table \ref{SuppTable:26Regions}). 
Since these spatial partitions sometimes required 
arbitrary subdivisions (e.g. breaking up USA, China, Japan and Australia), we also applied a discrete location scheme that accommodated a single location for these spatially and administratively coherent regions, arriving at 15 geographic regions  (Table \ref{SuppTable:15Regions}).
Within each sampling year, we randomly down-sampled the five locations with the highest number of samples relative to their population size (USA: from 278 to 150; Australia: from 166 to 30;  New Zealand: from 59 to 20; Japan: from  341 to 75; South Korea: from 51 to 30) and analyzed three different subsampled data sets.

\newcommand{\mymark}[1]{$^{#1}$}
\newcommand{\myfootnote}[2]{%
{\hspace*{0.5in}\tiny{$^{#1}$ #2} \\
\vspace*{-0.1in}%
}%
}

\begin{table}[h]
\caption{\footnotesize{\textbf{Absolute latitude, population size, population density, H3N2 sequence sample sizes and antigenic residuals for 26 global geographic regions.}
Antigenic residual estimates are described in \ref{GLMpredictors}.}}
\begin{center}
\vspace*{-0.25in}
\resizebox{1\textwidth}{!}{
\begin{tabular}{lccccc}
\hline\hline
Region &
Absolute latitude &
Population size &
Population density &
H3N2 sample size &
Antigenic residual \\
 & 
(degrees)  & & 
(people per km$^2$)  & & \\
\hline
Africa\mymark{1} & 14.03 & 1.96E+08 & 26.67 & 10 & -1.18\\
Canada\mymark{2} & 50.67 & 3.23E+07 & 2.82 & 27 & 0.18\\
Europe & 50.34 & 4.84E+08 & 96.18 & 80 & -0.51\\
Indochina\mymark{3} & 13.83 & 7.96E+07 & 114.72 & 59 & -0.30\\
New Zealand & 42.03 & 4.03E+06 & 14.89 & 59 & 0.07\\
Russia\mymark{4} & 53.36 & 1.47E+08 & 7.86 & 24 & -0.92\\
East China & 28.30 & 3.06E+08 & 453.29 & 56 & 0.57\\
Mexico & 19.43 & 1.03E+08 & 52.49 & 10 & 0.19\\
Mideast Japan & 35.71 & 4.54E+07 & 895.36 & 86 & -0.03\\
Midwest China & 29.82 & 3.08E+08 & 159.11 & 42 & 1.44\\
Midwest Japan & 35.18 & 3.52E+07 & 573.15 & 80 & -0.13\\
North China & 40.27 & 1.24E+08 & 308.49 & 33 & 0.91\\
Northeast Australia & 25.35 & 4.60E+06 & 2.42 & 79 & -0.91\\
Northeast Japan & 39.43 & 1.77E+07 & 108.60 & 82 & -0.19\\
Northeast USA\mymark{5} & 41.15 & 5.45E+07 & 129.59 & 48 & 0.64\\
Northwest Australia & 24.72 & 2.15E+06 & 0.55 & 27 & 0.17\\
West USA\mymark{5} & 40.34 & 3.20E+07 & 12.13 & 74 & 0.19\\
South America & 15.73 & 3.12E+08 & 20.71 & 57 & -0.57\\
West \& South Asia\mymark{6} & 27.17 & 1.28E+09 & 369.15 & 20 & -2.33\\
South Australia & 36.77 & 1.35E+07 & 6.48 & 60 & -0.74\\
South China & 22.33 & 1.02E+08 & 563.84 & 50 & 0.89\\
South Korea & 36.17 & 4.73E+07 & 475.35 & 52 & -0.08\\
Southeast Asia\mymark{8} & 5.47 & 1.19E+08 & 169.04 & 58 & 0.19\\
South USA\mymark{5} & 33.61 & 1.16E+08 & 51.59 & 104 & 0.50\\
Southwest Japan & 34.02 & 2.51E+07 & 275.43 & 93 & -0.36\\
Midwest USA\mymark{5} & 42.02 & 5.30E+07 & 44.78 & 62 & 0.28\\
\end{tabular}
}
\end{center}
\vspace*{-0.25in}
\myfootnote{}{}
\myfootnote{1}{includes Algeria, Egypt, Madagascar, South Africa and Saudi Arabia}
\myfootnote{2}{includes Canada and Alaska}
\myfootnote{3}{includes Cambodia and Thailand}
\myfootnote{4}{includes Russia and Mongolia}
\myfootnote{5}{USA is partitioned according to the US census bureau regions}
\myfootnote{6}{includes India, Nepal and Bangladesh}
\myfootnote{7}{includes Philippines, Singapore,  Malaysia and Guam}
\label{SuppTable:26Regions}
\end{table}


\begin{table}
\caption{\footnotesize{\textbf{Absolute latitude, population size, population density, H3N2 sequence sample and sub-sample sizes and antigenic residuals for 15 global geographic regions.}}}
\begin{center}
\vspace*{-0.25in}
\resizebox{1.0\textwidth}{!}{
\begin{tabular}{lccccc}
\hline\hline
Region &  Absolute latitude & Population size & Population density & H3N2 sample size & Antigenic residual \\
 & (degrees)  &   & (people per km$^2$)  & & \\
\hline
Africa & 14.03 & 1.96E+08 & 26.67 & 10 & -1.18\\
Canada & 50.67 & 3.23E+07 & 2.82 & 27 & 0.18\\
Europe & 50.34 & 4.84E+08 & 96.18 & 80 & -0.51\\
Indochina & 13.83 & 7.96E+07 & 114.72 & 59 & -0.30\\
New Zealand & 42.03 & 4.03E+06 & 14.89 & 59/20\mymark{1} & 0.07\\
Russia & 53.36 & 1.47E+08 & 7.86 & 24 & -0.92\\
China & 29.19 & 8.40E+08 & 263.06 & 181 & 0.92\\
Japan & 36.01 & 1.23E+08 & 336.98 & 341/75\mymark{1} & -0.18\\
Australia & 29.38 & 2.02E+07 & 2.57 & 166/30\mymark{1} & -0.70\\
USA & 38.35 & 2.56E+08 & 39.37 & 278/150\mymark{1} & 0.39\\
Mexico & 19.43 & 1.03E+08 & 52.49 & 10 & 0.19\\
South America & 15.73 & 3.12E+08 & 20.71 & 58 & -0.57\\
South Asia & 27.17 & 1.28E+09 & 369.15 & 20 & -2.33\\
South Korea & 36.17 & 4.73E+07 & 475.35 & 51/30\mymark{1} & -0.08\\
Southeast Asia & 5.47 & 1.19E+08 & 169.04 & 58 & 0.19\\
\end{tabular}
}
\end{center}
\vspace*{-0.25in}
\myfootnote{}{}
\myfootnote{1}{Sample sizes are provided before/after down-sampling}
\label{SuppTable:15Regions}
\end{table}

Because passenger flux emerged as the main predictor in our phylogeographic model (see \ref{GLM}), we also identified discrete air communities in the worldwide air transportation network (see \ref{PassengerFlux}) and applied these as location states to our sequence sample.
To increase sequence numbers for under-sampled air communities, we also complemented the hemagglutinin gene sequences with publicly available sequences from Africa ($n=21$), USA (Hawaii, $n= 4$), Central America ($n = 13$), South America ($n = 46$) and Canada ($n=10$).
From this data set, we removed six sequences that appeared to be outliers in a root-to-tip divergence versus sampling time regression analysis, resulting in a total of 1529 sequences.
Within each sampling year, we randomly down-sampled the four locations with the highest number of samples relative to their population size (USA: from 318 to 120; Oceania: from 225 to 50;  Japan: from  327 to 75; Southeast Asia: from 175 to 100) and analyzed three different subsampled data sets discretized according to the 14 air communities.

\begin{table}[H]
\caption{\footnotesize{\textbf{Absolute latitude, population size, population density, H3N2 sequence sample and sub-sample sizes and antigenic residuals for 14 global air communities.}}}
\begin{center}
\vspace*{-0.25in}
\resizebox{1.0\textwidth}{!}{
\begin{tabular}{lccccc}
\hline\hline
Region &  Absolute latitude & Population size & Population density & H3N2 sample size & Antigenic residual \\
 & (degrees)  &   & (people per km$^2$)  & & \\
\hline
Africa & 22.27 & 8.03E+07 & 44.53 & 23 & -0.93\\
USA & 37.33 & 2.95E+08 & 32.21 & 318/120\mymark{1}  & 0.34\\
Taiwan & 25.04 & 2.28E+07 & 629.54 & 17 & 0.25\\
China & 32.42 & 1.29E+09 & 134.88 & 122 & 0.88\\
Russia & 55.60 & 1.44E+08 & 8.44 & 17 & -1.05\\
Oceania & 32.69 & 2.46E+07 & 3.06 & 225/50\mymark{1}  & -0.51\\
West \& South Asia & 27.61 & 1.38E+09 & 208.34 & 26 & -0.11\\
Japan & 36.03 & 1.28E+08 & 342.62 & 327/75\mymark{1}  & -0.22\\
Mexico & 19.65 & 1.03E+08 & 52.49 & 12 & 0.46\\
South America & 18.02 & 3.12E+08 & 20.71 & 101 & -0.32\\
Canada & 48.53 & 3.16E+07 & 3.17 & 24 & 0.11\\
Europe & 48.44 & 4.84E+08 & 96.18 & 85 & -0.58\\
Southeast Asia & 15.27 & 2.06E+08 & 69.71 & 175/100\mymark{1}  & 0.10\\
South Korea & 36.01 & 4.73E+07 & 475.35 & 46 & -0.04\\
\end{tabular}}
\end{center}
\vspace*{-0.25in}
\myfootnote{}{}
\myfootnote{1}{Sample sizes are provided before/after down-sampling}
\label{SuppTable:14Communities}
\end{table}

\subsection{Air transportation data and identification of air transportation communities}

\subsubsection{Passenger flux data}\label{PassengerFlux}

The worldwide air transportation network is defined by a passenger flux matrix that quantifies the number of passengers traveling between each pair of airports.
We use a dataset provided by OAG (Official Airline Guide) Ltd. (\url{http://www.oag.com}), containing 4,092 airports and the number of seats on scheduled commercial flights between pairs of airports during the years 2004-2006.
The number of seats on scheduled commercial flights from airport $i$ to $j$ is given by \seats, which we take to be proportional to the number of passengers traveling.
For the 26 location scheme, we summarized the number of passengers from the full aviation network for each pair of locations based on all the airports in the respective regions.
To facilitate the identification of 14 air communities, we focused on a 1227-largest-airport network that represented 95\% of the passenger flux of the full aviation network by excluding the lowest contributing airports. By focusing on this subset of largest airports, we exclude a large number of very small community airports; details and plausibility of this reduction are discussed in \cite{woolley-meza12}.

\subsubsection{Modularity maximization}\label{MM} 

To identify air transportation communities, we approximate a maximal-modularity subdivision of the 1227-largest-airport network by employing a recently described stochastic Monte-Carlo approach \cite{thiemann10},
a generalization of the method introduced in \cite{guimera05}. 
Modularity provides a measure of how well the connectivity of a network is described by partitioning its nodes into non-overlapping groups.
For any given partition, modularity will be high if connectivity within groups is high and connectivity among groups is low.
In large networks, it is generally computationally impossible to find the optimal subdivision.
To approximate the optimum a variety of approximative methods have been introduced.
The method we employ here generates an ensemble of high modularity subdivisions and computes the consensus in this ensemble by superposition, for details see \cite{thiemann10,grady12}.


For an ensemble of 1000 modularity subdivisions we quantify the uncertainty by an affinity matrix that, for each pair of locations, summarizes the fraction of partitions in which these locations are in the same community.
Based on a partition encompassing a reasonably large number of air communities ($n$ = 14), we subsequently obtain the average affinity for each airport to the communities in this partition.
We assign each airport to the community for which it shows the highest average affinity, but we take into account its uncertainty by also considering assignments that yield affinities that are $>$ 2/3 of the highest affinity score.
This cut-off resulted in 771 ambiguous airport assignments.
Finally, we partitioned the sequence data according to the air community assignment and accommodate 368 (24\%) ambiguous sequence locations, i.e. those sequences related to airports with ambiguous community assignments, using ambiguity coding in our phylogeographic approach.

\subsection{Bayesian statistical analysis of sequence and trait evolution}

We integrate genetic, 
spatial 
and air transportation
data within a single full probabilistic evolutionary model and simultaneously estimate the parameters of phylogeographic 
diffusion using Markov chain Monte Carlo (MCMC) analysis implemented in BEAST \cite{drummond12}.
We introduce novel models  and inference procedures in the section below.
To model sequence evolution, we partition the hemagglutinin codon positions into first+second and third positions \cite{shapiro06} and apply a separate HKY85 \cite{ha85} CTMC model of nucleotide substitution with discrete gamma-distributed rate variation \cite{ya95} to both. We assume a flexible Bayesian skyride prior over the unknown phylogeny \cite{minin08a}.
Exploratory runs using the data for the 26 locations indicated that a relaxed molecular clock represented an over-parametrization \cite{drummond06}.
A strict clock was therefore used in subsequent analyses.
Because the exact date of sampling was not known for some additional publicly available sequences, 
we integrated out their dates over the known sampling time interval \cite{shapiro11}.
We capitalize on BEAGLE \cite{suchard09} in conjunction with BEAST to improve computational performance on our large data sets.
MCMC analyses were run sufficiently long to ensure stationarity as diagnosed using Tracer.
We used the TreeAnnotator tool in BEAST to summarize trees in the form of maximum clade credibility (MCC) trees.

\subsubsection{Phylogeographic inference and hypothesis testing}\label{GLM}

We develop a novel model-based approach to simultaneously reconstruct spatiotemporal history and test the contribution of potential predictors of spatial diffusion.
This approach builds on recently developed Bayesian phylogeographic inference methods to simultaneously reconstruct phylogenetic history and discretized diffusion processes \cite{lemey09}.
These processes are modeled as continuous-time Markov chain processes parameterized in terms of a $K$ x $K$ infinitesimal rate matrix of discrete location change (\rateMatrix).
Here, we extend this model by adopting a generalized linear model (GLM) approach that considers every rate of movement (\rateMatrixElement) in \rateMatrixSpace as a log linear function of an arbitrary number of predictors $\bf X$, such that:
\begin{equation}
log \rateMatrixElement = \coefficient_{1}\indicator_{1}log(\predictor_{1}) + 
\coefficient_{2}\indicator_{2}log(\predictor_{2}) + ... +
\coefficient_{n}\indicator_{n}log(\predictor_{n}),
\end{equation}
for \textit{n} predictors, where \coefficientSpace represents the effective size for the predictor log \predictor, quantifying its contribution to \rateMatrixElement, and \indicatorSpace is an (0,1)-indicator variable that governs the inclusion or exclusion of the predictor in the model.
The incorporation of indicator variables allows
 Bayesian stochastic search variable selection (BSSVS) \cite{kuo98,chipman01},
 which estimates
  the posterior probabilities of all possible linear models that may or may not include the predictors.
When an indicator equals 1, this predictor is included in the model, demonstrating that it helps to explain the diffusion process in the phylogenetic history with high probability.
We complete this GLM specification with variable selection by assigning independent Bernoulli prior probability distributions on \indicator, effectively placing equal probability on each predictorÕs inclusion and exclusion.
Lemey \textit{et al.} (2009)
\cite{lemey09}
discuss BSSVS in further detail and analogous to
Edo-matas \textit{et al.} (2011)
\cite{edomatas11}, we can use Bayes factors (BFs) \cite{jeffreys98,suchard01} to express how much the data change our prior opinion about the inclusion of each predictor.
These BFs are calculated by dividing the posterior odds for the inclusion of a predictor with the corresponding prior odds (here, 1:1 odds). 
\begin{equation}
	\mathrm{BF_i} = \frac{p_i}{1-p_i} / \frac{q_i}{1-q_i},
	\label{BF-equation}
\end{equation}  
where $p_i$ is the posterior probability that the predictor is included, in this case the posterior expectation of indicator $\indicator_i$, and $q_i$ is the prior probability that $\indicator_i = 1$. 
We specify that \textit{a priori} the \coefficient's are normally distributed with mean 0 and a variance of 4.
We implement the GLM-diffusion parametrization in the software package BEAST \cite{drummond2012bayesian} and approximate the joint posterior and its marginalizations using standard Markov chain Monte Carlo (MCMC) transition kernels.
An important, novel extension to the standard MCMC machinery in BEAST lies in generating an efficient Metropolis-Hastings proposal distribution for the GLM coefficients $\boldsymbol{\beta}$.  
Given the potential for high correlation between predictors $\bf X$, attempting to update one coefficient $\beta_j$ at a time while holding the remaining constant returns high autocorrelations times.  Instead, we exploit the fixed correlation structure ${\bf X}' {\bf X}$ between predictors to generate a multivariate proposal $\boldsymbol{\beta}^{\star}$. In particular, if we assume $\boldsymbol{\beta}$ are the current realized values, then we draw
\begin{equation}
    \boldsymbol{\beta}^{\star} \sim \text{Multivariate-Normal}
    \left(
    \boldsymbol{\beta}, \alpha 
    \left({\bf X}'{\bf X}\right)^{-1}
    \right),
\end{equation}
where $\alpha$ is an auto-tunable variance scalar.  
Motivation for this proposal stems from imagining that the marginal posterior distribution of $\boldsymbol{\beta}$ under our phylogenetic GLM should partially approximate a simple linear regression model involving $\boldsymbol{\beta}$, whose posterior variance is proportional to ${\bf X}'{\bf X}$.
We consider a `bit-flip' operator on the Bernoulli rate indicators; this transition kernel is further discussed in \cite{drummond10}.

We extended the phylogeographic inference techniques to take into account ambiguous location states in order to accommodate the uncertainty of the modularity maximization procedure in assigning airports to distinct discrete air communities (see \ref{MM}).

\subsubsection{GLM-diffusion predictors}\label{GLMpredictors}

Depending on the location state partitioning scheme, we considered several potential predictors of global influenza diffusion:

\begin{itemize}

\item \textbf{Average and minimum distance}.
To test whether geographical proximity predicts influenza diffusion we considered two different great-circle distance measures: (i) the average distance between two locations based on the pairwise distances between all pairs of airports from the two locations and (ii) the minimum distance amongst those pairwise distances.

\item \textbf{Absolute latitude}.
Absolute latitudes for each region/community were calculated as the absolute values of the average latitudes of the sequence sampling locations (Sequences from unknown locations within specific countries were assigned to the capital of that country) and are listed in Tables \ref{SuppTable:26Regions}, \ref{SuppTable:15Regions} and  \ref{SuppTable:14Communities}.

\item \textbf{Passenger flux}.
The total number of passengers traveling between each pair of locations per day (see \ref{PassengerFlux}).

\item \textbf{Population size and density}. Demographic estimates obtained from \url{www.citypopulation.de} or Geographica \cite{geographica} are listed in Tables \ref{SuppTable:26Regions},  \ref{SuppTable:15Regions} and  \ref{SuppTable:14Communities}. Origin and destination population sizes/densities were included as separate predictors.

\item \textbf{Viral surveillance data}.
To test the predictive power of viral surveillance data, we essentially aimed at capturing the nature and degree of synchronicity of yearly incidence profiles in each region. 
To this purpose, we extracted the number of influenza viruses A(H3) detected per country from week 1 in 1997 to week 45 in 2010 from FluNet/WHO (\url{www.who.int/flunet}) for relevant countries in the discrete partition schemes.
Taiwanese surveillance data was obtained from
\cite{lin11}.
We focused on the influenza A(H3) incidence counts between 2002 - 2007 or as close as possible to this time period when insufficient data was available.
Average incidence counts were used when data from multiple countries per region/community was available.
We subsequently calculated average incidences per week across multiple years for each region/community, normalized these weekly averages and smoothed them with a Gaussian standard deviation of 2 weeks. 
Figure \ref{SuppFigure:Incidence} depicts the resulting incidence profiles for the 14 air communities.

\begin{figure}
  \begin{center}
    \includegraphics[width=0.9\textwidth]{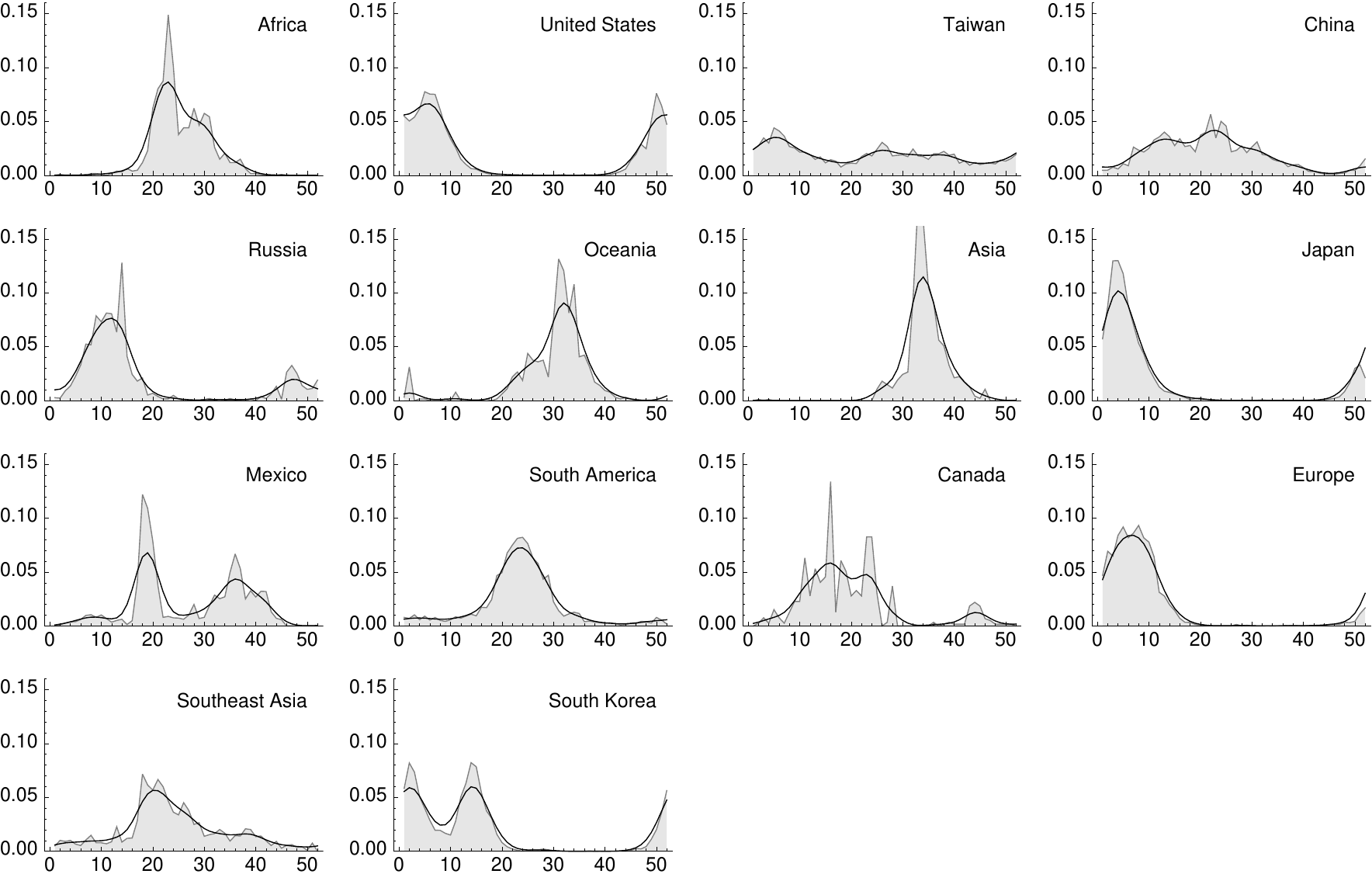}
  \end{center}
  \caption{\footnotesize{\textbf{H3 incidence profiles for the 14 air communities based on both raw and smoothed, normalized, weekly average counts for relevant countries in the air communities.} For the following regions we include data from specific countries: Africa (South Africa), Oceania (Australia), South \& West Asia (Bangladesh and India), South America (Argentina, Brazil, Chile, Ecuador, Paraguay, French Guiana, Peru, Uruguay and Venezuela), Europe (Bulgaria, Croatia, Czech Republic, Denmark, Finland, France, Germany, Greece, Iceland, Ireland, Italy, Latvia, Netherlands, Norway, Romania, Slovenia, Spain, Sweden, Switzerland, Turkey, Ukraine, UK and Ireland,) and Southeast Asia (Malaysia, Philippines, Singapore and Thailand).}}
\label{SuppFigure:Incidence}
\end{figure}

We derived several potential predictors from these incidence profiles, including incidence overlap, origin incidence versus destination growth rate, peak time difference, and incidence in the origin location at fixed times prior to peak incidence in the destination location.
The incidence overlap summarizes the overlapping area under the origin-destination incidence curves for each pair of locations. 
The origin incidence versus destination growth rate sums the product of origin incidence and destination growth rate for each week of the year.
Peak time difference quantifies the difference in peak incidence for each origin-destination pair.
For the latter, we summarized the donor incidences at 4, 8, 12, 16, 20 and 24 weeks prior to peak incidence in the destination location as potential predictors.

\item \textbf{Antigenic evolution}.
Because antigenic evolution can provide insights into the seeding dynamics of seasonal H3N2 \cite{russell08}, we sought to include the average antigenic divergence for each region as phylogeographic diffusion predictors.
Based on the available antigenic cartography data for the strains in our phylogeographic analyses, we performed a local regression (LOESS) of the principle antigenic component, obtained from a multidimensional scaling analysis of hemagglutination inhibition assay measurements \cite{russell08}, against time.
The resulting scatter plot with strains colored according to air community is presented in Figure \ref{SuppFigure:AntigenicResiduals}.
Distances from the spline (residuals) were calculated for each antigenic measurement and average residuals were obtained for each region/community, which reflects whether a location is on average antigenically leading or trailing \cite{russell08}.
These average residuals are listed in Tables  \ref{SuppTable:26Regions}, \ref{SuppTable:15Regions} and \ref{SuppTable:14Communities}.
We considered the exponentiated residual and exponentiated negative residual as a measure of efflux and influx respectively for each location and included these as separate origin and destination predictors.

\begin{figure}
  \begin{center}
    \includegraphics[width=0.65\textwidth]{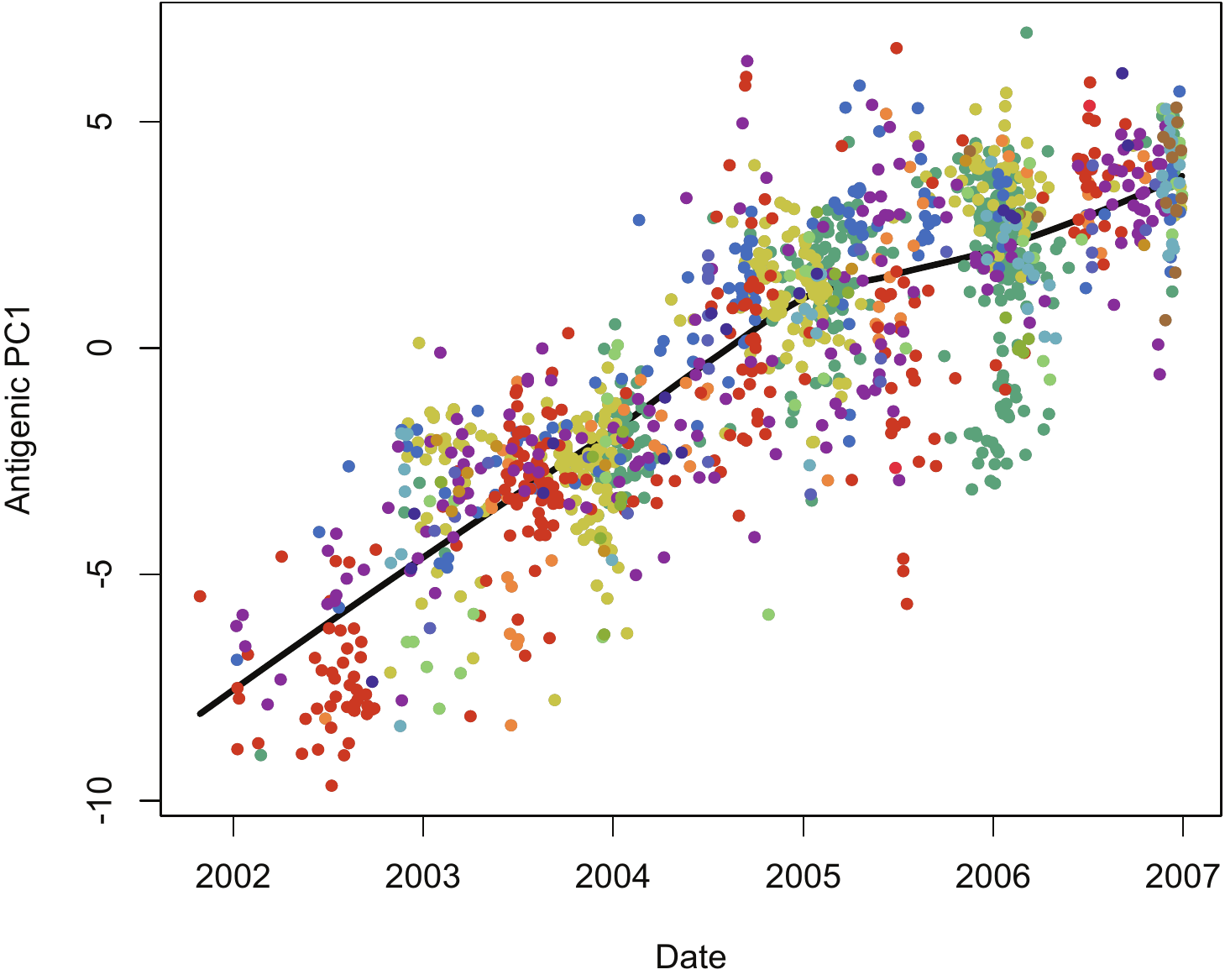}
  \end{center}
  \caption{\footnotesize{\textbf{Scatter plot with local regression (LOESS) fit for the first principle component (PC1) of the H3N2 antigenic measurements against time.} Antigenic PC1 data points are colored according to the air communities represented in Fig. \ref{SuppFig:14Communities}.}}
  \label{SuppFigure:AntigenicResiduals}
\end{figure}

\item \textbf{Sample sizes}.
To test the impact of sampling effects, we considered origin and destination sample sizes (number of H3N2 sequences included per discrete location state in the phylogeographic analysis) as separate predictors.
Although sampling sizes are expected to have an impact on the number of location transitions, support for other factors in addition to sampling size predictors may suggest that they are robust to potential sampling biases.
\end{itemize}

All predictors were transformed to log space and standardized prior to their incorporation in the GLM approach.

\subsubsection{Fitting diffusion models to empirical tree distributions} 

Although we generally desire to simultaneously reconstruct sequence and discrete/continuous trait evolution using our Bayesian statistical framework, integrating over tree-space becomes a computationally daunting task for a large number of taxa.
The main limiting factor in Bayesian MCMC analysis of evolutionary history is typically the efficiency with which topology proposals sample phylogenetic tree space \cite{lakner08}.
To side-step these limitations and reduce time to convergence, we seek to approximate phylogenetic uncertainty in our phylogeographic estimates in cases where sampling tree space needs to be performed repeatedly (e.g. when comparing different diffusion models).
To this purpose, we follow
\cite{pagel04} and implement proposal mechanisms to randomly draw from an empirical posterior distribution of trees, which, in our case, were solely inferred from sequence data.
Because the likelihood of a tree topology will largely be dominated by an informative sequence alignment compared to a single discrete (location) site, we expect such an empirical distribution to closely approximate the phylogenetic uncertainty in the joint inference approach.

\section{Results}

\begin{figure}
  \begin{center}
    \includegraphics[width=0.85\textwidth]{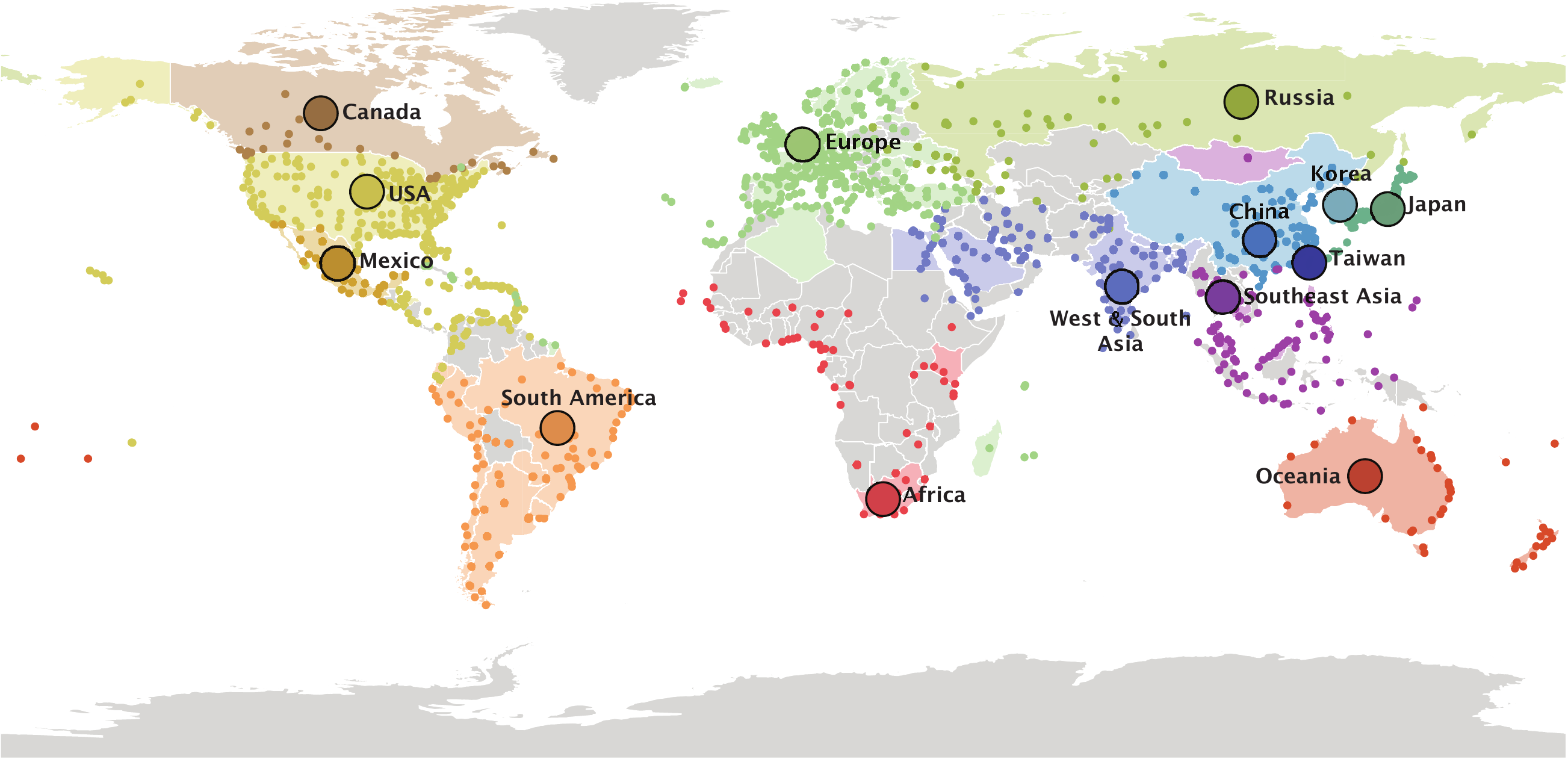}
  \end{center}
  \caption{\footnotesize{\textbf{14 global air communities identified through a modularity maximization analyses of air transportation data.}
  The colored dots represent the airports in each community for which passenger flux data was used in the analysis. 
  The areas with corresponding colors represent the geographical area within the communities for which H3N2 sequence samples were available.
  The 14 communities and associated data are listed in Table \ref{SuppTable:14Communities}.}}
\label{SuppFig:14Communities}
\end{figure}

To identify key factors in seasonal influenza dispersal, we inferred the phylogeographic history of globally sampled A/H3N2 viruses between 2002 and 2007, while simultaneously evaluating the contribution of several potential diffusion predictors using a novel Bayesian model selection procedure.
Our approach draws from recent developments in stochastic phylogenetic diffusion models \cite{lemey09}, and extends these by parameterizing pairwise diffusion rates as a function of a number of potential predictors, distinguishing between the 
statistical support for a predictor and the magnitude of its effect
 (see Methods).
We discretized the global sampling locations into 26 and 15 geographically defined regions (Table \ref{SuppTable:26Regions} and \ref{SuppTable:15Regions} respectively) as well as into 14 distinct air travel communities (Table \ref{SuppTable:14Communities}). 
To identify this community structure in global air travel, we determined partitions with high intra-module connectivity and low inter-module connectivity in a passenger flux network of 1227 airports.
Although this approach is blind to the geographic locations of the airports, the global air communities are spatially compact with few exceptions (Figure \ref{SuppFig:14Communities}).

\begin{figure}
  \begin{center}
    \includegraphics[width=1\textwidth]{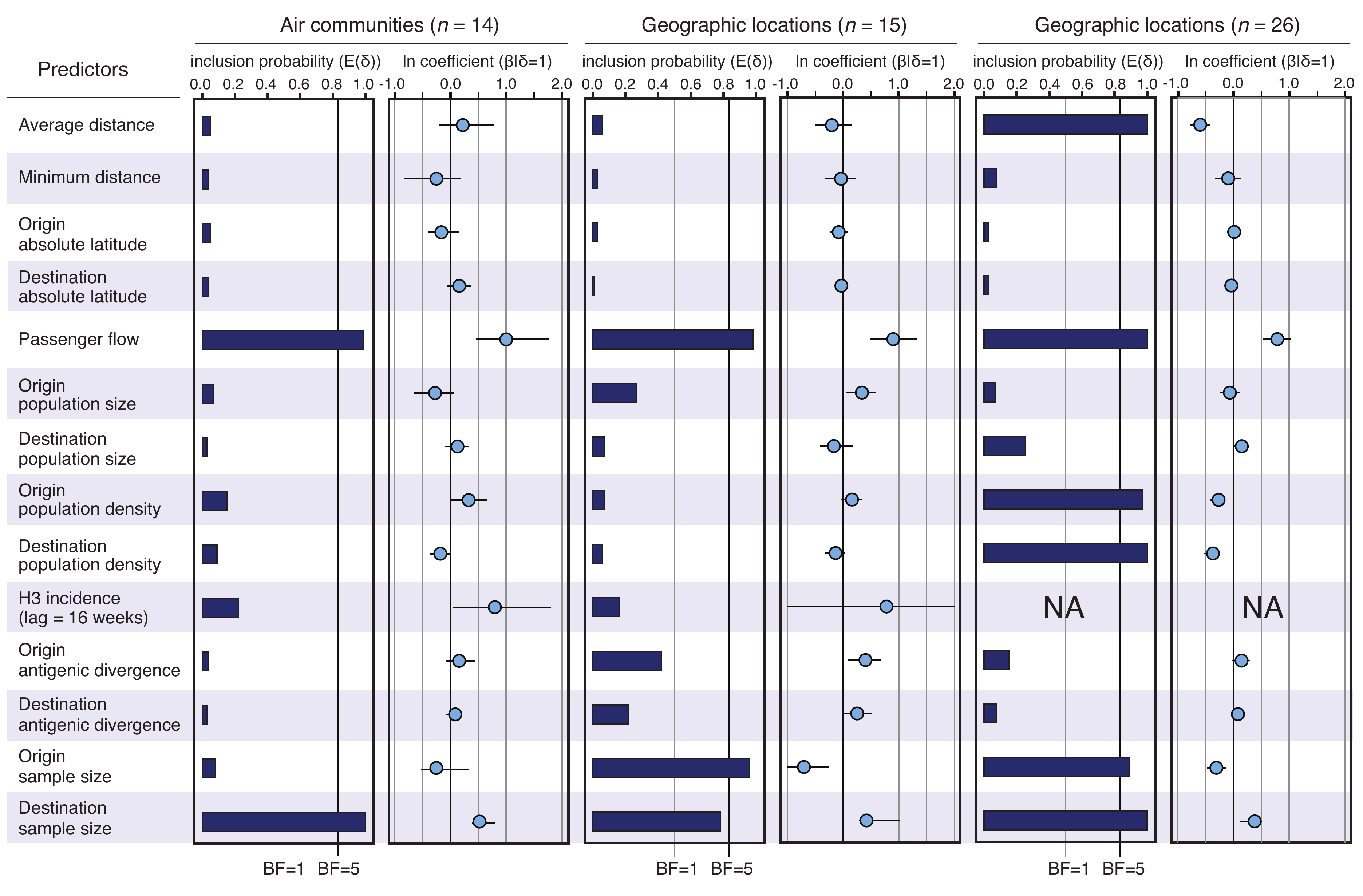}
  \end{center}
  \caption{\footnotesize{\textbf{Predictors of global H3N2 diffusion among the 14 air communities and the 15 \& 26 geographic locations.} The inclusion probabilities are defined by the indicator expectations $E[\delta]$ because they reflect the frequency by which the predictor is included in the model and therefore represent the support for the predictor. Bayes factor support values corresponding to two different indicator expectations, indicated a the bottom of the plots, are shown as vertical lines. The contribution of each predictor, when included in the model ($\beta|\delta=1$), is represented by the mean and credible intervals of the GLM coefficients on a log scale. }}
\label{Figure_predictors} 
\end{figure}

We compared a panel of possible predictors of phylogeographic diffusion using a generalized linear model (GLM) approach (see Methods). Here, we
considered geography, air travel (in the form of the number of passengers traveling between each pair of locations), demography, viral surveillance data, viral phenotypic evolution and sampling sizes as possible explanatory variables
(Figure \ref{Figure_predictors}).

Our phylogeographic test does not attribute any importance to geographical proximity, absolute latitude (to model source-sink behavior between the tropics and Northern/Southern hemisphere, \cite{rambaut08}), population size and density, antigenic divergence or H3 incidence for the 14 air communities and the 15 geographic regions.
Instead, we provide consistent and overwhelming evidence for passenger flow driving the global H3N2 diffusion dynamics, as reflected in the posterior support 
and a conditional effect size close to unity on a log scale for both the air community and geographic partitions.
Despite efforts to down-sample presumably oversampled regions or communities (see Methods), there is still a role for sample size in both the air community and geographic partitions.
Explicitly including sample sizes as diffusion predictors allows us to absorb potential effects of sampling bias, offering more credibility for other predictors that are included in the model.
When applied to 26 geographic regions, which further partitions geographically and administratively coherent regions like US, China, Japan and Australia (Table \ref{SuppTable:26Regions}), our model also takes an important negative contribution from geographic distance and, correlated with this but with lower coefficients, population densities. 
Here, distance most likely represents the role of other human mobility processes such as commuting, which has been shown to play a key role in the spread of influenza in the US \cite{viboud06}.
The negative population density effect may suggest that commuting is to some extent less likely to occur out of or into dense subpopulations.

\section{Summary}

Influenza prevention and control critically relies on our understanding of its geographical transmission patterns.
Here, we demonstrate the ability to jointly reconstruct phylogeographic history while identifying the factors that contribute epidemic spread from viral genetic data.
Our analysis of global influenza transmission provides evidence for the key role of air travel, which is highly intuitive and has long been predicted by modeling studies (e.g.
\cite{longini86}
), but remained difficult to ascertain from empirical data.
The predictors of influenza diffusion will undoubtedly be scale-dependent as indicated by the role of geographic distance within more confined geographic areas (Figure \ref{Figure_predictors}), and this may represent other forms of human mobility such as commuting \cite{viboud06}, which can be tested in future applications.
More generally, our novel phylogenetic diffusion approach may be applied to different infectious diseases problems and provide entirely new opportunities for testing how host ecology shapes the distribution of pathogen genetic data.

\section{Acknowledgements}
We thank Guandi Li for Matlab assistance, Hsin-Fu for providing the Taiwanese H3 incidence data and Jessica Hedge for collecting the coordinates for the H3N2 sequences. 
The research leading to these results has received funding from the European Union Seventh Framework Programme [FP7/2007-2013] under Grant Agreement no. 278433-PREDEMICS and ERC Grant agreement no. 260864.
MAS is in part supported by NSF grant DMS 0856099 and NIH grants R01 GM086887 and R01 HG006139.
Collaboration between MAS, AR and PL was supported by the National Evolutionary Synthesis Center (NESCent), NSF EF-0423641.
This work was supported by the Wellcome Trust [092807] to AR.
TB is supported by the Howard Hughes Medical Institute and the European Molecular Biology Organization.
DJS and CAR acknowledge EU FP7 programs EMPERIE (223498) and ANTIGONE (278976), Human Frontier Science Program (HFSP) program grant
P0050/2008, Wellcome 087982AIA, and NIH Director's Pioneer Award DP1-OD000490-01.
CAR was supported by a University Research Fellowship from the Royal Society.
Further, research was partially completed while PL and MAS were visiting the Institute for Mathematical Sciences, National University of Singapore in 2011.

\newpage

\newpage

\bibliographystyle{Science}


 \end{document}